\documentclass[aps,twocolumn,superscriptaddress,prl,showpacs]{revtex4}
\usepackage{graphicx}
\def\etal{~\emph{et al.}}
\begin{document}


\title{Optical alignment and polarization conversion of neutral exciton spin  in individual InAs/GaAs quantum dots}


\author{ K. Kowalik} \affiliation{CNRS-Laboratoire de Photonique et de Nanostructures,
Route de Nozay, 91460 Marcoussis, France} \affiliation{Institute of Experimental Physics,
University of Warsaw, Ho\.{z}a 69, 00-681 Warszawa, Poland}

\author{O.~Krebs} \email[Corresponding author :]{Olivier.Krebs@lpn.cnrs.fr}

\author{A.~Lema\^{i}tre}
\affiliation{CNRS-Laboratoire de Photonique et de Nanostructures, Route de Nozay, 91460
Marcoussis, France}

\author{J.~A.~Gaj} \affiliation{Institute of Experimental Physics, University of Warsaw, Ho\.{z}a 69, 00-681
Warszawa, Poland}

\author{P.~Voisin}
\affiliation{CNRS-Laboratoire de Photonique et de Nanostructures, Route de Nozay, 91460
Marcoussis, France}


\date{\today}

\begin{abstract}
We investigate exciton spin memory in individual InAs/GaAs self-assembled quantum dots via
optical alignment and conversion of exciton polarization in a magnetic field. Quasiresonant
phonon-assisted excitation is successfully employed to define the initial spin polarization of
neutral excitons. The conservation of the linear polarization generated along the bright exciton
eigenaxes of up to 90\%  and the conversion from circular- to linear polarization of up to 47\%
both demonstrate a very long spin relaxation time with respect to the radiative lifetime.
Results are quantitatively compared with a model of  pseudo-spin 1/2 including  heavy-to-light hole mixing.\\
\end{abstract}
\pacs{72.25.Fe, 72.25.Rb, 78.67.Hc,73.63.Kv}

\maketitle

\begin{figure}[h]
\begin{center}
\includegraphics[width=0.45\textwidth,keepaspectratio]{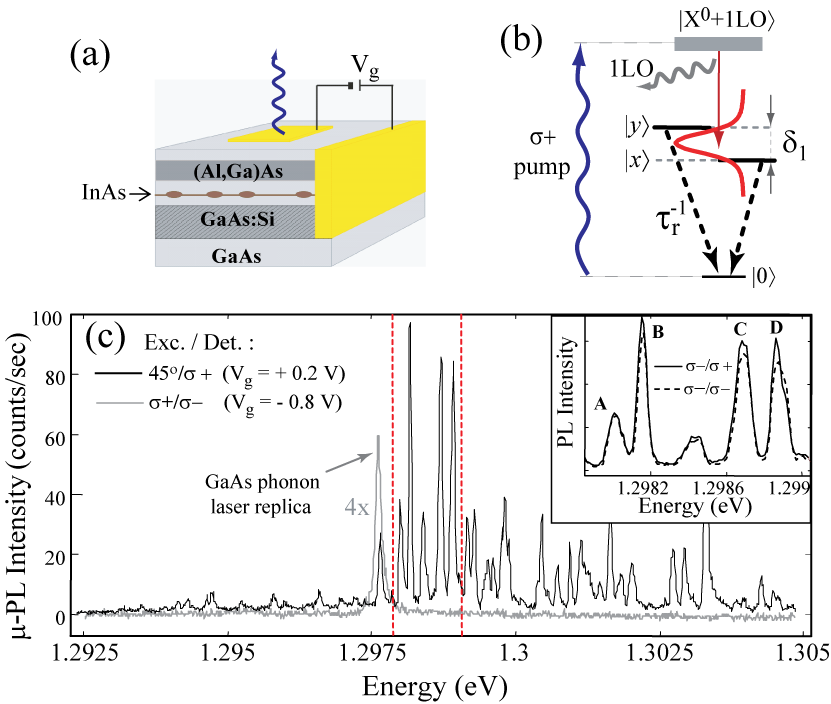}%
\end{center}
\caption{(a) Sample structure of InAs/GaAs QDs used for controlling exciton  charge state by
an electric field. (b) Scheme of 1~LO-phonon-assisted excitation. (c) $\mu$-PL spectra at
quasi-resonant excitation for two different voltages. Dark line -- neutral excitons observed
at +0.2~V ($45^o$ excitation, $\sigma +$ detection), gray line -- bulk GaAs Raman scattering
obtained at -0.8~V in $\sigma +$/$\sigma -$ configuration. Inset: zoom of the region defined
by the  dashed lines bringing out  4 excitonic lines (A, B, C, D) with vanishing circular
polarization under $\sigma^+$ excitation.}\label{fig:1}
\end{figure}

\begin{figure}[h]
\begin{center}
\includegraphics[width=0.45 \textwidth,keepaspectratio]{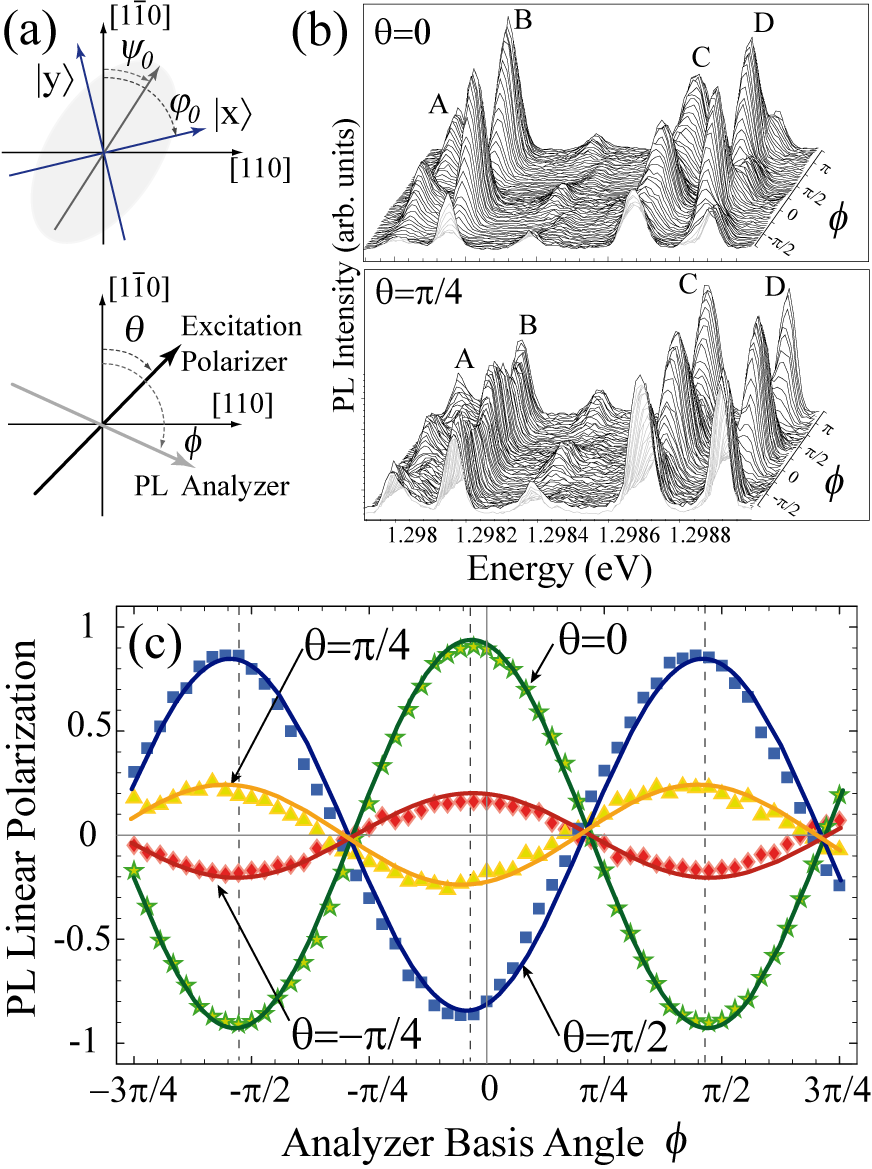}%
\end{center}
\caption{ (a) Scheme of the QD eigenaxis $\varphi_0$  and dichroism $\psi_0$ angles (top) and
settings of polarizers (bottom) with respect to the $[1\bar{1}0]$ direction. (b) $\mu$-PL
spectra measured for different PL linear analyzer angles $\phi$  and for two angles $\theta$
of the excitation polarization. (c) Linear polarization $P_l(\phi)$ of exciton B (see text),
for different linear polarizations $\theta$ of the excitation. Solid lines are theoretical fit
made with the same set of parameters.}\label{fig:2OK}
\end{figure}

\indent Since the first proposal of using nanostructures for quantum
computation~\cite{Loss-DiVincenzo} the semiconductor quantum dots (QDs) have been attracting a
lot of attention. In such a system, quantum  information can be stored and manipulated by
using the degrees of freedom of a single or a few charge carriers, provided the coupling to
the environment does not destroy the coherence. Single spin manipulation in a QD has been
recently achieved by electron spin resonance~\cite{Koppens-Nature,Kroner-ESR} and
time-resolved Faraday or Kerr
rotation~\cite{Atatuere-NPhys3,Mikkelsen-NPhys3,Sci313-Greilich}, evidencing unequivocally the
long coherence time of electron spin. In contrast, the spin coherence of a dipole-active
neutral exciton (an electron hole pair) has been recently questioned  in
Ref.~\cite{PRB-Favero} : from rather indirect observations the authors conclude  about a fast
exciton spin relaxation ($<$100~ps) in self-assembled InAs/GaAs quantum dots.  In this system,
the bright excitons form  a two-level system which can be considered as a pseudo-spin 1/2. In
an ideal (cylindrical) QD , the eigenstates $|\pm1\rangle$ are coupled to  $\sigma^{\pm}$
circularly polarized photons, but in actual QDs some symmetry reduction generally lifts the
exciton degeneracy between  two linearly-polarized states $|x\rangle$ and $|y\rangle$ by the
fine structure splitting $\delta_1\equiv\hbar \Omega_{exc.}$ (see Fig.~\ref{fig:1}(b)). Yet,
the polarization quenching of these states was  already demonstrated for excitons in an
ensemble of InAs QDs under resonant excitation~\cite{Paillard-PRL86}.  Investigating the
longitudinal relaxation of single exciton spin by a straightforward method is thus highly
demanded to
elucidate the issue raised by Ref.~\cite{PRB-Favero}.\\
 \indent In this work, we report on  close to 100\% optical alignment of
the exciton dipole in individual quantum dots and close to 50\% (theoretical maximum)
conversion of polarization from \emph{circular} to \emph{linear} in a longitudinal magnetic
field. These observations demonstrate that the exciton spin relaxation is  quite negligible on
the timescale of the exciton lifetime ($\sim$1~ns)  for all the quantization directions
defined by the effective magnetic field experienced by
the quantum dot.\\
\indent The sample was grown by molecular beam epitaxy on a semi-insulating GaAs [001]
substrate. It contains  a vertical field effect structure with a single layer of InAs/GaAs
quantum dots (see Fig.~\ref{fig:1}(a)). The detailed description of the structure can be found
in Ref.~\cite{PRL94-Laurent}. Photoluminescence (PL) of  neutral excitons was observed in the
bias range [-0.1~V,0.25~V] so that for all the experiments reported here, the voltage was fixed
at  +0.2~V. The PL excitation and collection were performed through the semitransparent part of
the top Schottky gate.\\
\indent The $\mu$-PL spectroscopy of individual InAs QDs was performed using a split-coil
magneto-optical cryostat. A 2~mm focal length aspheric lens (N.A.~0.5) was used to focus the
excitation beam from a cw Ti:Sapphire laser and to collect the PL from the sample, while the
relative positioning in all three directions was ensured by Attocube$^{TM}$ piezo-motors with a
precision of $\sim$0.5~$\mu$m. This very compact microscope  was immersed in the pumped liquid
helium bath of the cryostat insert. All measurements were performed at low temperature (T=2~K)
and the magnetic field was applied parallel to the optical axis (Faraday configuration). The PL
was dispersed using a 0.6~m-focal length \emph{double} spectrometer and detected using a
Nitrogen-cooled CCD array camera. A Babinet-Soleil compensator was used to correct for the
residual birefringence of the cube beam splitter which separates the excitation and collection
beams. The resulting extinction ratio was below 0.01 for both circular and linear
cross-polarizations~\cite{PhD-Kowalik}.\\
\indent   A prerequisite for the study of the exciton spin relaxation in QDs, is the ability
to create a coherent superposition  or  a certain mixed state of the bright exciton states
$|\pm 1\rangle$. To achieve this we used the quasiresonant 1~LO-phonon-assisted excitation
(see Fig.~\ref{fig:1}(b)) which is compatible with standard cw PL spectroscopy of individual
QDs. In InAs/GaAs quantum dots a marked resonance in the excitation spectrum  occurs at $\sim
35$~meV above the ground state transition, i.e. between the GaAs LO- and TO-phonon
features~\cite{PRL94-Laurent,Oulton-PRB68,Lemaitre-PRB63,Ignatiev-PRB63}. Since the electron
phonon interaction is essentially spin independent, this technique enables in principle to
initialize the exciton spin state as performed in quantum wells~\cite{Ivchenko-conversion1},
or more recently in self-assembled CdSe/ZnSe
QDs~\cite{Flissikowski-PRL86,Astakhov-circ2linConversion,Kusrayev-PRB72} and InAs/GaAs
QDs~\cite{Senes-PRB71}. Creating a coherent state under these conditions requires that the
exciton splitting $\delta_1$  be not much greater than the exciton-phonon (or polaron)
linewidth. Since the latter is determined by the fast decay (below 10 ps) of the LO-phonon
part into acoustical phonons~\cite{Verzelen-PRL88} this generally occurs, and indeed coherent
exciton states have been evidenced by quantum beat measurements both in CdSe quantum
dots~\cite{Flissikowski-PRL86} and InAs quantum dots~\cite{Senes-PRB71}. We therefore consider
in the following  such a coherent
state is generated, although our cw experiments did not enable us  to probe the exciton coherence.\\
\indent As illustrated in Fig.~\ref{fig:1}(c), the weak LO-phonon Raman scattering of bulk GaAs
could be identified at exactly $\hbar\omega_{LO}$=36.6~meV below the laser energy in the
cross-circular configuration of the excitation and detection polarizations and in an electric
field of $F_z\sim$100~kV/cm such that the PL signal from QDs is completely suppressed. At a
positive bias (0.2~V) a few lines corresponding to neutral excitons are observed above the GaAs
phonon replica of the laser which is likely due to  the coupling with LO-phonons from InGaAs
alloy of slightly lower energy~\cite{Oulton-PRB68}. In the rest of the paper we focus on $4$
specific exciton lines labelled A,B,C, and D chosen in a 1~meV wide spectral range  after the
laser was finely tuned to optimize their average intensity. Table~\ref{Table-FSS} gives the
exciton parameters relevant to the study of the exciton spin, namely the fine structure
splitting $\delta_1$ and the angle $\varphi_0$ of the low-energy eigenaxis (chosen as the
$|x\rangle$ state, see Fig.~\ref{fig:2OK}(a)), measured by analyzing the linear polarization of
the PL under circularly polarized excitation. All excitons exhibit a typical FSS of a few tens
of $\mu$eV with eigenaxes which are significantly tilted from the $\langle110\rangle$
crystallographic axes. It is therefore crucial to investigate the properties of a single QD to
avoid averaging effects making the analysis less convincing~\cite{Kusrayev-PRB72}. Since these
splittings are much larger than the exciton homogeneous linewidth $\hbar/\tau_r <1$~$\mu$eV, the
optical orientation of excitons photocreated with $\sigma^+$ polarization averages to zero due
to quantum beats~\cite{Paillard-PRL86} as shown in the inset of Fig.~\ref{fig:1}(c). This
contrasts with the case of charged
excitons~\cite{PRL94-Laurent,Sci312-Atatuere}.\\
\begin{table}
\begin{ruledtabular}
\begin{tabular}{ccccc}
Exciton & A & B & C & D \\
FSS $\delta_1$~($\mu$eV) & 26 & 16 & 29 & 50 \\
Eigenaxis angle $\varphi_0$~($^\circ$)  & 69 & 83 & 121 & 155 \\
\end{tabular}\end{ruledtabular}
\centering \caption{Parameters of fine structure splitting of  all four selected
excitons.}\label{Table-FSS}
\end{table}
\indent  We first consider the optical alignment of  excitons. The excitation linear
polarization was set parallel ($\theta=0,\pi/2$) or at 45$^\circ$ ($\theta=\pm\pi/4$) to the
$\langle110\rangle$ axes in order to create an exciton either in one of its eigenstates
$|x\rangle$, $|y\rangle$ or as a coherent superposition $(|x\rangle\pm|y\rangle)/\sqrt{2}$. As
the QD eigenaxes were found tilted from the $\langle110\rangle$ directions (see
Tab.~\ref{Table-FSS}) our choice of polarizer directions was not optimal  but nevertheless
sufficient to reveal strong exciton dipole alignment. The PL linear polarization was analyzed
with a polarizer making an angle $\phi$ with the $[1\bar{1}0]$ direction (see
Fig.~\ref{fig:2OK}(a)). Two series of spectra obtained by varying  $\phi$ are presented in
Fig.~\ref{fig:2OK}(b) for two settings of the polarizer, and  the degree of linear
polarization of exciton B is plotted in Fig.~\ref{fig:2OK}(c). All excitonic lines exhibit
strong intensity oscillations as a function of the analyzer angle, with maxima and minima
corresponding to their eigenaxes, and, remarkably with  an amplitude which depends on the
excitation polarization. For exciton B which has its eigenaxes tilted  by only 7$^\circ$ from
the $\langle110\rangle$ axes, this effect is particularly pronounced. The oscillation contrast
reaches $\sim$90\% for $\theta=0,\pi/2$, namely when the exciton dipole is almost oriented
along a QD eigenaxis, but falls down below $\sim$20\% when the initial exciton is aligned at
45$^\circ$ ($\theta=\pm\pi/4$). This demonstrates a very efficient optical alignment that can
be quantitatively discussed by plotting the degree of linear polarization
$P_l(\phi)=(I_{\phi}-I_{\phi+\pi/2})/(I_{\phi}+I_{\phi+\pi/2})$ where $I_{\phi}$ is the
integrated PL intensity of the exciton line for the analyzer at $\phi$ (see
Fig.~\ref{fig:2OK}(c)). By treating the bright exciton as a  pseudo-spin 1/2 experiencing an
effective in-plane magnetic field
$\Omega_{exc.}=\delta_1/\hbar$~\cite{Ivchenko-conversion2,Paillard-PRL86}, we can
straightforwardly derive in the density matrix formalism an analytical expression for
$P_l(\phi)$:
\begin{equation}
P_l(\phi)=\eta P_l^0\frac{(2+\nu^2)\cos 2(\phi-\theta)+\nu^2\cos
2(\phi+\theta-2\varphi_0)}{2(1+\nu^2)} \label{eq:OptAl}
\end{equation}
\noindent where $P_l^0$ is the photo-generated initial polarization,
$\eta=\tau_s/(\tau_s+\tau_r)$ with $\tau_s$ the exciton spin relaxation
time~\cite{pure-dephasing} and $\tau_r$ the exciton lifetime, and $\nu=\eta
\Omega_{exc.}\tau_r$. In Eq.~\ref{eq:OptAl} we have neglected the effect of exciton
thermalization within the bright doublet which would contribute an additional
term~\cite{thermalization}. Indeed, as a result of the observed large degree of polarization
we necessarily have $\eta>0.92$ so that the thermalization term contributes less than 1\% at
T=2~K. Yet, to explain the small asymmetry of the polarization amplitude between $\theta=0$
(0.91) and $\theta=\pi/2$ (0.87), as well as a small shift of polarization maxima for
$\theta=\pm\pi/4$ it is necessary to include linear dichroism of the bright exciton due to the
reduced symmetry of the QD, which has been modeled by assuming heavy-hole to light-hole
mixing~\cite{Koudinov-PRB70,Kowalik-PRB75,Leger-PRB76,Toropov-PRB63}. The $|\pm1\rangle$
bright exciton states which become elliptically polarized are written as
$|\pm1\rangle=\sqrt{1-\beta^2}|\mp\frac{1}{2},\pm\frac{3}{2}\rangle+\beta
e^{\pm2i\psi_0}|\mp\frac{ 1}{2},\mp\frac{ 1}{2}\rangle$ with $\beta>0$ the fraction of
light-hole exciton and $\psi_0$ the angle of dichroism main axis with respect to $[1\bar{1}0]$
(see Fig.~\ref{fig:2OK}(a)). As recently discussed for  CdTe QDs~\cite{Leger-PRB76}, this
angle may differ greatly from the exchange eigenaxis orientation $\phi_0$.  The resulting
changes of optical selection rules give an additional term in Eq.~\ref{eq:OptAl} which to the
first order in $\beta$ can be written:
\begin{equation}
\Delta P_l(\phi)=-\frac{2}{\sqrt{3}}\beta \cos 2(\phi-\psi_0)\label{eq:HH-HL}
\end{equation}
This theoretical model provides a very good agreement with the experimental results  as shown
for exciton B in Fig.~\ref{fig:2OK}(c) by taking $\beta=0.04$, $\psi_0=93^\circ$,
$P_l^0=0.95$, $\tau_r$=0.85~ns, $\tau_s=$20~ns. We can therefore conclude that (i) the exciton
spin can be optically written with a very high fidelity under phonon assisted excitation and
(ii) the exciton spin relaxation time is at least one order of magnitude longer than its
radiative
lifetime.\\
\begin{figure}[h]
\begin{center}
\includegraphics[width=0.45\textwidth,keepaspectratio]{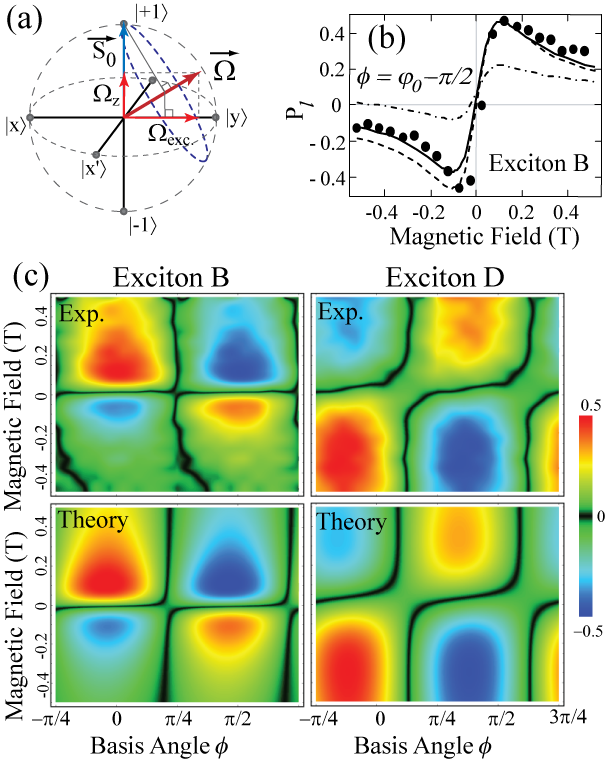}%
\end{center}
\caption{(a) Scheme  of the exciton spin precession on the Poincarr\'{e} sphere in presence of an
external magnetic field $\Omega_z$. (b) C2L conversion of polarization for exciton B.
Experimental points (dots), theoretical model  (solid line: $\tau_s=10 \tau_r$, $\beta=0.04$,
dashed line: $\tau_s=10 \tau_r$, $\beta=0$, dot-dashed line: $\tau_s= 0.5 \tau_r$,
$\beta=0.04$). (c) Contour-plot of the degree of linear polarization as a function of the
magnetic field and
 angle $\phi$ of analysis for excitons B and D (top), and theoretical simulation (bottom). See
text for details. }\label{fig:3}
\end{figure}
\indent To investigate further the exciton  spin relaxation along  directions different from the
QD eigenaxes, a magnetic field $B_z$ was applied parallel to the optical axis.  The latter is
described below by the  corresponding Larmor precession angular frequency $\Omega_z= g_{exc.}
\mu_B B_z/\hbar$, where $g_{exc.}$ is the exciton $g$~factor and $\mu_B$ the Bohr magneton.  The
total field $\vec{\Omega}=(\Omega_{exc.},0,\Omega_z)$, about which the initial exciton spin
$\vec{S}_0$ precesses, becomes oblique with respect to both the optical axis and the QD
eigenaxes as sketched in the Poincarr\'{e}'s sphere representation (see Fig.~\ref{fig:3}(a)). Since
the average exciton spin acquires a finite projection on the $|x\rangle$ or $|y\rangle$ states,
this gives rise to the phenomenon of polarization conversion previously studied in QWs or QD
ensembles~\cite{Ivchenko-conversion1,Ivchenko-conversion2,Kusrayev-PRB72,Astakhov-circ2linConversion,Paillard-PRL86}.
For an individual QD, this effect is spectacular because the broadening due to inhomogeneous
distribution of anisotropic splittings $\delta_1$ is suppressed. Note however that polarization
conversion would still occur in case of a mixed state (e.g. resulting from fast decoherence of
the spin component perpendicular to $\vec{\Omega}$) as long as the relaxation of the spin
component parallel to $\vec{\Omega}$ is negligible. The exciton coherence really manifests
itself in cw experiments only through the polarization conversion in the direction orthogonal
both to $\vec{\Omega}$ and $\vec{S}_0$ (sine term in Eq.~\ref{eq:C2L} given below), which is
however much weaker
because of the multiple  periods of precession during the  exciton lifetime~\cite{Astakhov-circ2linConversion}.\\
\indent Experimentally we have observed both linear-to-circular (L2C) and circular-to-linear
(C2L) conversion for all the excitonic lines in Fig.~\ref{fig:1}. Here we focus  on the C2L
conversion which has been investigated by measuring the linear polarization of individual
lines for different angles $\phi$ of the analyzer under a fixed $\sigma^+$ polarized
excitation, while the magnetic field was varied from -0.5~T to +0.5~T with a  50~mT step. The
results for exciton B and D are displayed  in Fig.~\ref{fig:3}. C2L conversion manifests
itself by the development of a strong linear polarization along the QD eigenaxes
($\phi=\varphi_0\,[\pi/2]$), which depends almost antisymmetrically on $B_z$ and presents
clear maxima for  finite values of the field (at about $\pm$100~mT for exciton B, see
Fig.~\ref{fig:3}(b)). Remarkably, this maximum reaches a value 0.47 which is close to the
theoretical upper limit (0.5) achieved when $\Omega_z=\Omega_{exc.}$ so that the precession
vector $\vec{\Omega}$ lies at 45$^\circ$ between $|+1\rangle$ and $|x\rangle$ spin states. For
a quantitative analysis, we calculated the field-induced degree of linear polarization for an
arbitrary angle $\phi$ by solving the density matrix evolution of the exciton spin, and
obtained the following analytical expression:
\begin{equation}
P_{l}(\phi)= \sqrt{1-\gamma^2}\frac{\eta \nu P_c^0}{1+\nu^2}\left[\sin
2(\phi-\varphi_0)-\gamma\nu\cos2(\phi-\varphi_0) \right] \label{eq:C2L}
\end{equation}
\noindent where $P_c^0$ is the photo-generated initial circular polarization,
$\gamma=\Omega_z/\Omega$, and now $\nu=\eta \Omega\tau_r$. As for optical alignment
experiments, the spin thermalization could be neglected. The observed departure from a perfect
anti-symmetrical dependence on $B_z$ (predicted by Eq.~\ref{eq:C2L}) can again be explained by
the linear dichroism of the excitons
given by Eq.~\ref{eq:HH-HL} as shown in Fig.~\ref{fig:3}(b). 
For an overall comparison with the model, the linear polarization as a function of magnetic
field and analysis angle $\phi$ has been plotted using a color scale (see
Fig.~\ref{fig:3}(c)). Note for a better rendering (``only'' 19 angles and 21 field values were
measured) an interpolation has been performed. These 2D plots reveal clearly how the basis of
maximum C2L conversion rotates from $\phi=\varphi_0+\pi/4\,[\pi/2]$ to
$\phi=\varphi_0\,[\pi/2]$, when the magnetic field strength increases, in agreement with both
righthand terms in Eq.~\ref{eq:C2L}. By taking into account the exciton parameters of
Tab.~\ref{Table-FSS}, we could reproduce reasonably well the experimental contour-plots with
Eqs.~\ref{eq:HH-HL},\ref{eq:C2L} as illustrated in the bottom part of Fig.~\ref{fig:3}(c). For
both excitons B and D, we used $g_{exc.}=2.5$, $\tau_r=0.85$~ns, $\tau_s=10$~ns, but different
values of valence band mixing, namely $\beta=0.04$ (0.1) and
$\psi_0=93^\circ$ $(90^\circ)$ for exciton B (D) respectively.\\ 
\indent In summary,  optical alignment of exciton spin along each of its linearly polarized
eigenstates was demonstrated in zero field, whereas a very efficient conversion of polarization
from a circularly- to linearly-polarized exciton was achieved by applying a longitudinal
magnetic field. Our results, discussed within the framework of a theoretical model including
valence band mixing effects on exciton optical selection rules, prove straightforwardly that the
exciton spin of an isolated QD has a very long relaxation time in comparison to the exciton
radiative lifetime, a conclusion which is of fundamental importance
for the development of QD-based sources emitting polarization entangled photon pairs.\\
\indent One of us (K.K.) was supported by the European network of excellence SANDIE and by the
Foundation for Polish Science (START programme). This work has been partially supported by the
Polish Ministry of Science and Higher Education (Grants financed in 2007-2010).

\begin{thebibliography}{27}
\expandafter\ifx\csname natexlab\endcsname\relax\def\natexlab#1{#1}\fi \expandafter\ifx\csname
bibnamefont\endcsname\relax
  \def\bibnamefont#1{#1}\fi
\expandafter\ifx\csname bibfnamefont\endcsname\relax
  \def\bibfnamefont#1{#1}\fi
\expandafter\ifx\csname citenamefont\endcsname\relax
  \def\citenamefont#1{#1}\fi
\expandafter\ifx\csname url\endcsname\relax
  \def\url#1{\texttt{#1}}\fi
\expandafter\ifx\csname urlprefix\endcsname\relax\def\urlprefix{URL }\fi
\providecommand{\bibinfo}[2]{#2} \providecommand{\eprint}[2][]{\url{#2}}

\bibitem[{\citenamefont{Loss and DiVincenzo}(1998)}]{Loss-DiVincenzo}
\bibinfo{author}{\bibfnamefont{D.}~\bibnamefont{Loss}} \bibnamefont{and}
  \bibinfo{author}{\bibfnamefont{D.~P.} \bibnamefont{DiVincenzo}},
  \bibinfo{journal}{Phys. Rev. A} \textbf{\bibinfo{volume}{57}},
  \bibinfo{pages}{120} (\bibinfo{year}{1998}).

\bibitem[{\citenamefont{Koppens\etal}(2004)}]{Koppens-Nature}
\bibinfo{author}{\bibfnamefont{F.~H.} \bibnamefont{Koppens\etal}},
  \bibinfo{journal}{Nature} \textbf{\bibinfo{volume}{442}},
  \bibinfo{pages}{766} (\bibinfo{year}{2004}).

\bibitem[{\citenamefont{Kroner\etal}(2007)}]{Kroner-ESR}
\bibinfo{author}{\bibfnamefont{M.}~\bibnamefont{Kroner\etal}},
  \bibinfo{journal}{cond-mat arXiv:0710.4901v1}  (\bibinfo{year}{2007}).

\bibitem[{\citenamefont{Atat\"{u}re et~al.}(2007)\citenamefont{Atat\"{u}re,
  Dreiser, Badolato, and Imamoglu}}]{Atatuere-NPhys3}
\bibinfo{author}{\bibfnamefont{M.}~\bibnamefont{Atat\"{u}re}},
  \bibinfo{author}{\bibfnamefont{J.}~\bibnamefont{Dreiser}},
  \bibinfo{author}{\bibfnamefont{A.}~\bibnamefont{Badolato}}, \bibnamefont{and}
  \bibinfo{author}{\bibfnamefont{A.}~\bibnamefont{Imamoglu}},
  \bibinfo{journal}{Nature Phys.} \textbf{\bibinfo{volume}{3}},
  \bibinfo{pages}{521} (\bibinfo{year}{2007}).

\bibitem[{\citenamefont{Greilich\etal}(2006)}]{Sci313-Greilich}
\bibinfo{author}{\bibfnamefont{A.}~\bibnamefont{Greilich\etal}},
  \bibinfo{journal}{Science} \textbf{\bibinfo{volume}{313}},
  \bibinfo{pages}{341} (\bibinfo{year}{2006}).

\bibitem[{\citenamefont{Mikkelsen\etal}(2007)}]{Mikkelsen-NPhys3}
\bibinfo{author}{\bibfnamefont{M.~H.} \bibnamefont{Mikkelsen\etal}},
  \bibinfo{journal}{Nature Phys.} \textbf{\bibinfo{volume}{3}},
  \bibinfo{pages}{736} (\bibinfo{year}{2007}).

\bibitem[{\citenamefont{Favero\etal}(2005)}]{PRB-Favero}
\bibinfo{author}{\bibfnamefont{I.}~\bibnamefont{Favero\etal}},
  \bibinfo{journal}{Phys. Rev. B} \textbf{\bibinfo{volume}{71}},
  \bibinfo{pages}{233304} (\bibinfo{year}{2005}).

\bibitem[{\citenamefont{Paillard\etal}(2001)}]{Paillard-PRL86}
\bibinfo{author}{\bibfnamefont{M.}~\bibnamefont{Paillard\etal}},
  \bibinfo{journal}{Phys. Rev. Lett.} \textbf{\bibinfo{volume}{86}},
  \bibinfo{pages}{1634} (\bibinfo{year}{2001}).

\bibitem[{\citenamefont{Laurent\etal}(2005)}]{PRL94-Laurent}
\bibinfo{author}{\bibfnamefont{S.}~\bibnamefont{Laurent\etal}},
  \bibinfo{journal}{Phys. Rev. Lett.} \textbf{\bibinfo{volume}{94}},
  \bibinfo{pages}{147401} (\bibinfo{year}{2005}).

\bibitem[{\citenamefont{Kowalik}(2007)}]{PhD-Kowalik}
\bibinfo{author}{\bibfnamefont{K.}~\bibnamefont{Kowalik}}, Ph.D. thesis,
  \bibinfo{school}{Universit\'{e} Paris VI and Warsaw University}
  (\bibinfo{year}{2007}).

\bibitem[{\citenamefont{Oulton\etal}(2003)}]{Oulton-PRB68}
\bibinfo{author}{\bibfnamefont{R.}~\bibnamefont{Oulton\etal}},
  \bibinfo{journal}{Phys. Rev. B} \textbf{\bibinfo{volume}{68}},
  \bibinfo{pages}{235301} (\bibinfo{year}{2003}).

\bibitem[{\citenamefont{Lema\^{\i}tre\etal}(2001)}]{Lemaitre-PRB63}
\bibinfo{author}{\bibfnamefont{A.}~\bibnamefont{Lema\^{\i}tre\etal}},
  \bibinfo{journal}{Phys. Rev. B} \textbf{\bibinfo{volume}{63}},
  \bibinfo{pages}{161309(R)} (\bibinfo{year}{2001}).

\bibitem[{\citenamefont{Ignatiev\etal}(2001)}]{Ignatiev-PRB63}
\bibinfo{author}{\bibfnamefont{I.~V.} \bibnamefont{Ignatiev\etal}},
  \bibinfo{journal}{Phys. Rev. B} \textbf{\bibinfo{volume}{63}},
  \bibinfo{pages}{75316} (\bibinfo{year}{2001}).

\bibitem[{\citenamefont{Ivchenko\etal}(1991)}]{Ivchenko-conversion1}
\bibinfo{author}{\bibfnamefont{E.~L.} \bibnamefont{Ivchenko\etal}},
  \bibinfo{journal}{Superlattices and Microstr.} \textbf{\bibinfo{volume}{10}},
  \bibinfo{pages}{497} (\bibinfo{year}{1991}).

\bibitem[{\citenamefont{Flissikowski\etal}(2001)}]{Flissikowski-PRL86}
\bibinfo{author}{\bibfnamefont{T.}~\bibnamefont{Flissikowski\etal}},
  \bibinfo{journal}{Phys. Rev. Lett.} \textbf{\bibinfo{volume}{86}},
  \bibinfo{pages}{3172} (\bibinfo{year}{2001}).

\bibitem[{\citenamefont{Astakhov\etal}(2006)}]{Astakhov-circ2linConversion}
\bibinfo{author}{\bibfnamefont{G.~V.} \bibnamefont{Astakhov\etal}},
  \bibinfo{journal}{Phys. Rev. Lett.} \textbf{\bibinfo{volume}{96}},
  \bibinfo{pages}{27402} (\bibinfo{year}{2006}).

\bibitem[{\citenamefont{Kusrayev\etal}(2005)}]{Kusrayev-PRB72}
\bibinfo{author}{\bibfnamefont{Y.~G.} \bibnamefont{Kusrayev\etal}},
  \bibinfo{journal}{Phys. Rev. B} \textbf{\bibinfo{volume}{72}},
  \bibinfo{pages}{155301} (\bibinfo{year}{2005}).

\bibitem[{\citenamefont{S\'{e}n\`{e}s\etal}(2005)}]{Senes-PRB71}
\bibinfo{author}{\bibfnamefont{M.}~\bibnamefont{S\'{e}n\`{e}s\etal}},
  \bibinfo{journal}{Phys. Rev. B} \textbf{\bibinfo{volume}{71}},
  \bibinfo{pages}{115334} (\bibinfo{year}{2005}).

\bibitem[{\citenamefont{Verzelen et~al.}(2002)\citenamefont{Verzelen, Ferreira,
  and Bastard}}]{Verzelen-PRL88}
\bibinfo{author}{\bibfnamefont{O.}~\bibnamefont{Verzelen}},
  \bibinfo{author}{\bibfnamefont{R.}~\bibnamefont{Ferreira}}, \bibnamefont{and}
  \bibinfo{author}{\bibfnamefont{G.}~\bibnamefont{Bastard}},
  \bibinfo{journal}{Phys. Rev. Lett.} \textbf{\bibinfo{volume}{88}},
  \bibinfo{pages}{146803} (\bibinfo{year}{2002}).

\bibitem[{\citenamefont{Atat\"{u}re\etal}(2006)}]{Sci312-Atatuere}
\bibinfo{author}{\bibfnamefont{M.}~\bibnamefont{Atat\"{u}re\etal}},
  \bibinfo{journal}{Science} \textbf{\bibinfo{volume}{312}},
  \bibinfo{pages}{551} (\bibinfo{year}{2006}).

\bibitem[{\citenamefont{Dzhioev\etal}(1997)}]{Ivchenko-conversion2}
\bibinfo{author}{\bibfnamefont{R.~I.} \bibnamefont{Dzhioev\etal}},
  \bibinfo{journal}{Phys. Rev. B} \textbf{\bibinfo{volume}{56}},
  \bibinfo{pages}{13405} (\bibinfo{year}{1997}).

\bibitem[{pur()}]{pure-dephasing}
\bibinfo{note}{We assumed no pure dephasing of the exciton spin coherence, so
  that a single time $\tau_s$ is used.}

\bibitem[{the()}]{thermalization}
\bibinfo{note}{Linear polarization of PL due to thermalization within the
  exciton bright doublet is given by $\cos 2 (\phi-\varphi_0)
  (1-\eta)\tanh(\hbar\Omega_{exc.}/2k_B T)$ with $k_B$ the Boltzman constant.}

\bibitem[{\citenamefont{Koudinov et~al.}(2004)\citenamefont{Koudinov, Akimov,
  Kusrayev, and Henneberger}}]{Koudinov-PRB70}
\bibinfo{author}{\bibfnamefont{A.~V.} \bibnamefont{Koudinov}},
  \bibinfo{author}{\bibfnamefont{I.~A.} \bibnamefont{Akimov}},
  \bibinfo{author}{\bibfnamefont{Y.~G.} \bibnamefont{Kusrayev}},
  \bibnamefont{and}
  \bibinfo{author}{\bibfnamefont{F.}~\bibnamefont{Henneberger}},
  \bibinfo{journal}{Phys. Rev. B} \textbf{\bibinfo{volume}{70}},
  \bibinfo{pages}{241305(R)} (\bibinfo{year}{2004}).

\bibitem[{\citenamefont{Kowalik\etal}(2007)}]{Kowalik-PRB75}
\bibinfo{author}{\bibfnamefont{K.}~\bibnamefont{Kowalik\etal}},
  \bibinfo{journal}{Phys. Rev. B} \textbf{\bibinfo{volume}{75}},
  \bibinfo{pages}{195340} (\bibinfo{year}{2007}).

\bibitem[{\citenamefont{L\'{e}ger et~al.}(2007)\citenamefont{L\'{e}ger,
  Besombes, Maingault, and Mariette}}]{Leger-PRB76}
\bibinfo{author}{\bibfnamefont{Y.}~\bibnamefont{L\'{e}ger}},
  \bibinfo{author}{\bibfnamefont{L.}~\bibnamefont{Besombes}},
  \bibinfo{author}{\bibfnamefont{L.}~\bibnamefont{Maingault}},
  \bibnamefont{and} \bibinfo{author}{\bibfnamefont{H.}~\bibnamefont{Mariette}},
  \bibinfo{journal}{Phys. Rev. B} \textbf{\bibinfo{volume}{76}},
  \bibinfo{pages}{045331} (\bibinfo{year}{2007}).

\bibitem[{\citenamefont{Toropov\etal}(2000)}]{Toropov-PRB63}
\bibinfo{author}{\bibfnamefont{A.~A.} \bibnamefont{Toropov\etal}},
  \bibinfo{journal}{Phys. Rev. B} \textbf{\bibinfo{volume}{63}},
  \bibinfo{pages}{35302} (\bibinfo{year}{2000}).

\end{thebibliography}

\end{document}